Short Communication

# Limits to the power density of very large wind farms

Takafumi Nishino

*Department of Engineering Science, University of Oxford, Oxford OX1 3PJ, UK*

ABSTRACT

A simple analysis is presented concerning an upper limit of the power density (power per unit land area) of a very large wind farm located at the bottom of a fully developed boundary layer. The analysis suggests that the limit of the power density is about 0.38 times $\tau_{w0}U_{F0}$, where $\tau_{w0}$ is the natural shear stress on the ground (that is observed before constructing the wind farm) and $U_{F0}$ is the natural or undisturbed wind speed averaged across the height of the farm to be constructed. Importantly, this implies that the maximum extractable power from such a very large wind farm will not be proportional to the cubic of the wind speed at the farm height, or even the farm height itself, but be proportional to $U_{F0}$.

Correspondence:
Takafumi Nishino, Department of Engineering Science, University of Oxford, Parks Road, Oxford OX1 3PJ, UK. E-mail: takafumi.nishino@eng.ox.ac.uk

## 1. INTRODUCTION

With the development of the wind energy industry during the last few decades, research on the (prospective) performance of very large wind farms has been attracting increasing interests.[1-4] In general, the airflow within a large wind farm is considered to approach asymptotically to the so-called "fully developed" state as the horizontal extents of the farm become much larger (say, more than 10 to 20 times larger) than the thickness of the atmospheric boundary layer (ABL), which is about 1 km.

The characteristics of such a large wind farm consisting of a number of horizontal-axis wind turbines (HAWT's) have been investigated theoretically by e.g. Frandsen et al.[1,2] More recently, Calaf et al.[3] have performed large-eddy simulations (LES) of fully developed boundary layer flow over regularly aligned HAWT's and proposed a modification to the effective roughness model of Frandsen et al.[2] Meyers and Meneveau[4] have discussed an optimal turbine spacing in such fully developed wind farm boundary layers based on the results of Calaf et al.[3] Markfort et al.,[5] however, have performed wind tunnel tests of HAWT's in staggered as well as aligned configurations and reported that the characteristics of flow within a farm may depend significantly on the farm configuration.

Whilst many earlier studies have focused on wind farms consisting of common HAWT's, some other studies have considered using different types of wind turbines. For example, Dabiri[6] has recently reported field tests of 10-meter-tall vertical-axis wind turbines (VAWT's) in various counter-rotating configurations. The tests demonstrated that an array of counter-rotating VAWT's could achieve a much higher power density (power per unit land area) compared to that of existing wind farms consisting of HAWT's, although the array size tested was small relative to the ABL thickness and it has not been proven or demonstrated yet whether such a high power density can be achieved for large wind farms.

One fundamentally important question here is: *What is the (theoretical) upper limit of power density for very large wind farms?* Knowing such an upper limit would be useful, for example, when we assess the efficiency of land use for various types of wind farms. In this short communication, a simple theoretical analysis is presented to answer this question.

## 2. ANALYSIS

Let us consider a fully developed boundary layer forced by a constant streamwise pressure gradient. In the actual ABL, the flow is driven by pressure gradient and Coriolis forces, causing the so-called "Ekman spiral" in the outer (Ekman) layer of the ABL (see e.g. Tennekes and Lumley[7]). The pressure-driven boundary layer considered in this analysis, however, still serves as a good approximation to the actual ABL especially when the height of wind farms to be studied is much smaller than the thickness of the ABL, $\delta$, as discussed by e.g. Calaf et al.[3] We consider constructing a very large wind farm of uniform height $H_F$ (<< $\delta$) at the bottom of the fully developed boundary layer. We do not specify the type, number or array configuration of wind turbines to be deployed within the farm area, but assume that they are arrayed in a periodic manner (so that the entire farm area can be divided into a number of sub-areas, in each of which one or more turbines are placed in an identical manner). The horizontal extents of the farm area are assumed to be much larger than $\delta$, and hence the flow across the majority of the farm area will be fully developed, i.e. the flow profile horizontally averaged across each sub-area will not change across the majority of the farm area.

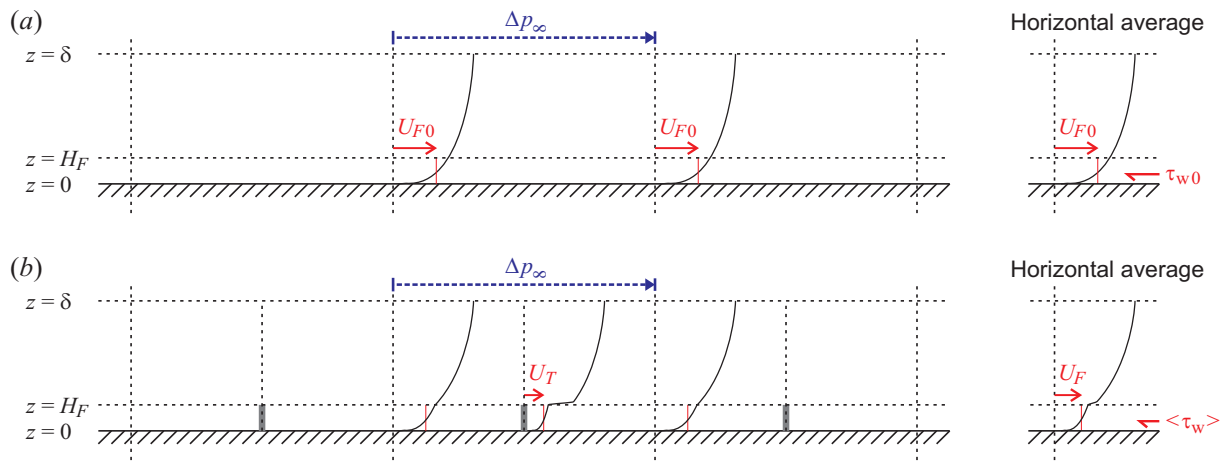

Fig. 1. Schematic of fully developed boundary layer flows across a small part of a very large wind farm site: (*a*) before and (*b*) after constructing the farm.

Figure 1 shows a schematic of two fully developed boundary layer flows across a small part (three sub-areas) of a very large wind farm site: one for before and the other for after the construction of the farm. Although the figure has been depicted two-dimensionally, the analysis to be presented is applicable to general three-dimensional cases, where turbines are arrayed periodically in both streamwise (*x*) and cross-stream (*y*) directions. Also remember that more than one turbine may be placed in each sub-area (of horizontal area *S*), even though only one turbine per sub-area is drawn in this figure for simplicity. The turbines in this figure are depicted such that they extend across the entire height of the farm ($0 \leq z \leq$ $H_F$) but in general they may occupy only part of the farm height. Since the flow is driven by a constant streamwise pressure gradient, the pressure difference (or drop) across each sub-area, $\Delta p_\infty$, does not change due to the construction of the farm.

First, we define $U$ and $\tau_w$ as the (time-averaged) streamwise velocity and shear stress on the ground, respectively, and $\langle U \rangle$ and $\langle \tau_w \rangle$ as the spatial (horizontal) average of $U$ and $\tau_w$ across each sub-area, respectively. We then define $U_F$ as the average of $\langle U \rangle$ across the farm height (i.e. $U_F$ is the mean wind speed within the farm):

$$U_F = \frac{\int_0^{H_F} \langle U \rangle \, \mathrm{d}z}{H_F} \tag{1}$$

We also define $U_{F0}$ and $\tau_{w0}$ as $U_F$ and $\langle \tau_w \rangle$ for the case without the farm (i.e. the original values of $U_F$ and $\langle \tau_w \rangle$ observed before constructing the farm).

Here we consider the momentum balance for a control volume of height $\delta$ and base area $S$ (corresponding to each sub-area of the farm). Since the velocity field is periodic in both streamwise ($x$) and cross-stream ($y$) directions and also the (time-averaged) shear stress on the top of the control volume ($z = \delta$) is considered to be zero (provided that $H_F \ll \delta$), the following relationship is always satisfied:

$$\Delta p_\infty A = \langle \tau_w \rangle S + T = \tau_{w0} S \tag{2}$$

where $A$ is the frontal area of the control volume (which is $\delta$ times the extent of the control volume in $y$ direction) and $T$ is the total thrust on the turbine(s) within the control volume. Hence, if we employ the so-called "actuator disk" concept and assume the power extracted from the turbine(s), $P$, to be the product of the thrust ($T$) and the mean wind speed across the frontal area of the turbine(s), $U_T$, then the power is calculated as

$$P = TU_T = \left( \tau_{w0} - \langle \tau_w \rangle \right) S U_T \tag{3}$$

It should be noted that the assumed relationship $P = TU_T$ is not exact when the wind profile across the turbine(s) is not uniform; however, the discrepancy arising from this assumption is expected to be small. Since the actuator disk concept itself is a theoretical approximation of power extraction from a real turbine, we do not consider the discrepancy arising from the above assumption in this analysis.

Equation (3) shows clearly that the amount of power extracted from the turbine(s) is large when: (i) $\tau_{w0} - \langle \tau_w \rangle$ is large (i.e. $\langle \tau_w \rangle$ is small compared to its original value observed before constructing the farm); and (ii) $U_T$ is large. In order for $\langle \tau_w \rangle$ to decrease, the mean wind speed within the farm, $U_F$, usually needs to decrease, whereas $U_T$ is not larger than $U_F$ in general; therefore, there is an optimal $U_F$ to maximise the power extracted from the turbine(s). To examine such an optimal $U_F$ for an ideal wind farm, here we consider that the ratio of $\langle \tau_w \rangle$ to $\tau_{w0}$ can be described as

$$\frac{\langle \tau_w \rangle}{\tau_{w0}} = \left(\frac{U_F}{U_{F0}}\right)^\alpha \tag{4}$$

The exponent $\alpha$ is expected to be close to 2, although its exact value depends on how the flow profile across the farm height is actually altered by the turbine(s). (Note that $\alpha = 2$ means that a nominal friction coefficient defined as $C_f^* = \langle \tau_w \rangle / \frac{1}{2} \rho U_F^2$, where $\rho$ is the air density, does not change depending on $U_F$). By substituting equation (4) into equation (3), we obtain

$$P = \tau_{w0} \left\{ 1 - \left(\frac{U_F}{U_{F0}}\right)^\alpha \right\} S U_T \tag{5}$$

If we consider an ideal wind farm situation where the power is extracted horizontally homogeneously within each sub-area, we can assume that the mean wind speed across the turbines ($U_T$) can be as large as the mean wind speed within the farm ($U_F$). For such an ideal wind farm with $U_T = U_F$, equation (5) can be re-written as

$$\frac{P_{\text{ideal}}}{\tau_{w0} U_{F0} S} = \frac{U_F}{U_{F0}} - \left(\frac{U_F}{U_{F0}}\right)^{\alpha+1} \tag{6}$$

where the subscript "ideal" indicates that this power is for an ideal wind farm with $U_T = U_F$. Figure 2 plots solutions of equation (6), showing the power density (non-dimensionalised by $\tau_{w0} U_{F0}$) of an ideal wind farm as a function of $U_F/U_{F0}$ and $\alpha$. As can be seen from the figure, the power density first increases but then decreases as $U_F/U_{F0}$ decreases from 1 to 0. For the case with $\alpha = 2$, which is expected to be a good approximation (especially for the homogeneous power extraction considered herein), the non-dimensional power density reaches a peak value of $\sqrt{1/3} - \left(\sqrt{1/3}\right)^3 \approx 0.385$ when $U_F/U_{F0} = \sqrt{1/3} \approx 0.577$.

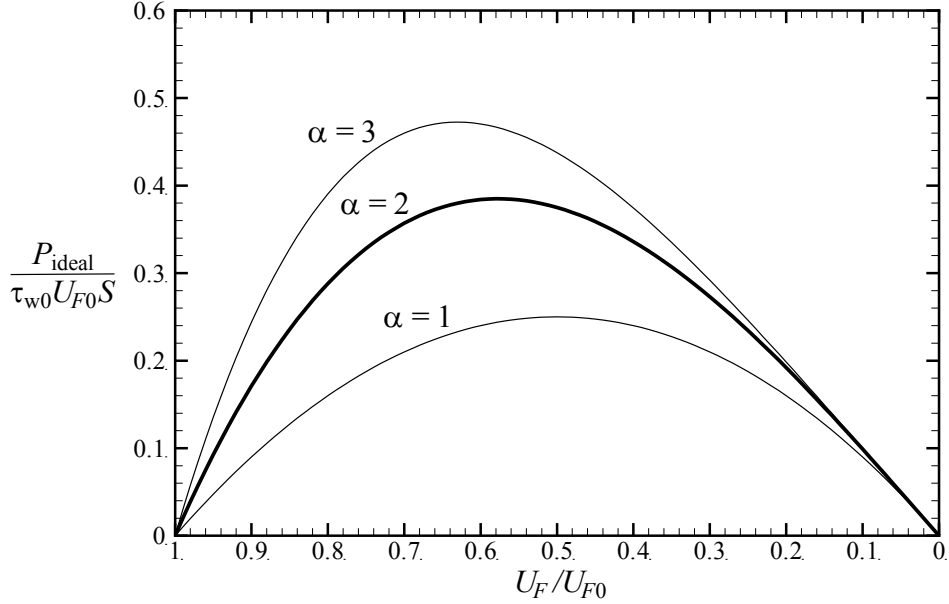

Fig. 2. Power density of an ideal wind farm as a function of $U_F/U_{F0}$ and $\alpha$.

## 3. CONCLUDING REMARKS

The present analysis has shown that the limit of the power density of a very large wind farm is about 0.38 times $\tau_{w0}U_{F0}$, provided that the nominal friction coefficient on the ground, defined as $C_f^* = \langle \tau_w \rangle / \frac{1}{2}\rho U_F^2$, does not change depending on $U_F$. Theoretically, this peak power density can be achieved when the mean wind speed within the farm is reduced to about 58% of its original value (by ideal turbine arrays extracting the power horizontally homogeneously across the farm). Most importantly, the analysis suggests that the maximum extractable power from such a very large wind farm will not be proportional to the cubic of the wind speed at the farm height, or even the farm height itself, but be proportional only to the first power of the (original) wind speed averaged across the farm height.

It should be remembered, however, that extracting wind power horizontally homogeneously (and thereby achieving $U_T = U_F$) is a very difficult challenge; in reality, $U_T$ would be smaller than $U_F$ and the power density would be lower than the ideal values shown in figure 2. Also note that, since the deployment of wind turbines usually increases the level of turbulence near the ground (and hence likely increases the value of $C_f^*$), the exponent $\alpha$ in equations (4) to (6) is likely to be smaller that 2 in practice. To maintain $C_f^*$ as low as possible and $\alpha$ as high as possible (perhaps by enhancing turbulent mixing only near the top of the farm and *not* near the ground) may therefore be another challenge for the design of very large wind farms. From these perspectives, dense arrays of vertical axis turbines do seem a promising option for very large wind farms, especially when the land available is limited and the power density of the farm, rather than the power per turbine, is of primary concern.